\documentclass[12pt]{article}
\usepackage{amsmath,amssymb,amsfonts,amsthm}

\title{Entropy squeezing and atomic inversion  in the $k$-photon Jaynes-Cummings model in the presence of Stark shift and Kerr medium: full nonlinear approach}

\author{H R Baghshahi$^{1,2,3}$, M K Tavassoly$^{1,2,*}$ and A Behjat$^{1,4}$ \\
 \footnotesize{$^1$ Atomic and Molecular Group, Faculty of Physics, Yazd University, Yazd, Iran} \\
 \footnotesize{$^2$ The Laboratory of Quantum Information Processing, Yazd University, Yazd, Iran} \\
 \footnotesize{$^3$ Department of Physics, Faculty of Science, Vali-e-Asr University of Rafsanjan, Rafsanjan, Iran} \\
 \footnotesize{$^4$ Photonics Research Group, Engineering Research Center, Yazd University, Yazd, Iran} \\
 \footnotesize{$^*$ E-mail: mktavassoly@yazd.ac.ir}}

\begin{document}
\maketitle

 \newcommand{\norm}[1]{\left\Vert#1\right\Vert}
 \newcommand{\abs}[1]{\left\vert#1\right\vert}
 \newcommand{\set}[1]{\left\{#1\right\}}
 \newcommand{\R}{\mathbb R}
 \newcommand{\I}{\mathbb{I}}
 \newcommand{\C}{\mathbb C}
 \newcommand{\eps}{\varepsilon}
 \newcommand{\To}{\longrightarrow}
 \newcommand{\BX}{\mathbf{B}(X)}
 \newcommand{\HH}{\mathfrak{H}}
 \newcommand{\A}{\mathcal{A}}
 \newcommand{\D}{\mathcal{D}}
 \newcommand{\N}{\mathcal{N}}
 \newcommand{\x}{\mathcal{x}}
 \newcommand{\p}{\mathcal{p}}
 \newcommand{\la}{\lambda}
 \newcommand{\af}{a^{ }_F}
 \newcommand{\afd}{a^\dag_F}
 \newcommand{\afy}{a^{ }_{F^{-1}}}
 \newcommand{\afdy}{a^\dag_{F^{-1}}}
 \newcommand{\fn}{\phi^{ }_n}
 \newcommand{\HD}{\hat{\mathcal{H}}}
 \newcommand{\HDD}{\mathcal{H}}

 \begin{abstract}
In this paper the interaction between a two-level atom and a single-mode field in the $k$-photon  Jaynes-Cummings model (JCM) in the presence of Stark shift and Kerr medium is studied. All terms in the respected Hamiltonian, such as the single-mode field, its interaction with the atom, the contribution of the Stark shift and the Kerr medium effects are considered to be $f$-deformed. In particular, the effect of the initial state of radiation field on the dynamical evolution of some physical properties such as atomic inversion and entropy squeezing are investigated  by considering different initial field states. To achieve this purpose, coherent, squeezed and thermal states as initial field states are considered.
 \end{abstract}


  \section{Introduction}\label{sec-intro}
  The well-known Jaynes-Cummings model (JCM) is an important, simplified and standard model that describes elegantly the interaction between an atom and a single-mode  field in the dipole and rotating wave approximations (RWA) \cite{Jaynes}. Many interesting physical features have been studied by this model. Some examples are atomic inversion \cite{HU}, collapse and revival \cite{Scully,Setare}, entanglement \cite{Buzek12,Tan,Ouyang}, sub-Poissonian statistics \cite{Mandel,Mirzaee}, quadrature squeezing \cite{Rui} and entropy squeezing \cite{Fang,Jian}. A lot of researches in this field are based on the linear interaction between atom and field, i.e. the atom-field coupling is performed to be a constant throughout the evolution of the whole system. Phoenix and Knight \cite {Phoenix} used the JCM and employed a diagonalised reduced density operator to calculate entropy and demonstrated thereby the essential two-state nature of the field. Kayham investigated the entropy squeezing of a two-level atom interacting with a quantum field prepared initially in the Glauber-Lachs state by the standard JCM \cite{Kayham}. Liao {\it et al} considered a system of two two-level atoms interacting with a binomial field in an ideal cavity and investigated the time evolution of the single-atom entropy squeezing, atomic inversion and linear entropy of the system \cite{Liao}. Zhang {\it et al} discussed the entanglement and  evolution of some of the nonclassicality features of the atom-field system  in a standard JCM with  squeezed vacuum and coherent state fields as initial field state \cite{Zhang}. Mortezapour {\it et al} studied the entanglement of dressed atom and its spontaneous emission in a three-level $\Lambda$-type closed-loop atomic system in multi-photon resonance condition and beyond it \cite{Mortezapour}. The entropy squeezing, atomic inversion and variance squeezing in the interaction between a two-level atom with a single mode cavity field via $k$-photon process have been investigated in \cite{Kang}. Ateto in \cite{Ateto} has been extended the JCM for combined the influences of  atomic motion, field-mode structure and Kerr-like medium and investigated the effects of them on the dynamics of entropy of entanglement of the cavity and atomic populations.

However, in recent years, researches have strongly focused on the nonlinear interaction between a two-level atom and field in the deformed JCM. This model which firstly suggested by Buck and Sukumar \cite{Buck,Sukumar} describes the dependence of atom-field coupling on the light intensity. Bu\v{z}ek investigated the physical quantities, particularly atomic population and squeezing in the intensity-dependent coupling JCM \cite{Buz^ek}. The interaction between a $\Lambda$-type three-level atom with a single-mode cavity field with intensity-dependent coupling in a Kerr medium has been investigated by one of us \cite{Faghihi}. Sanchez and R\'{e}camier introduced a nonlinear JCM constructed from its standard structure by deforming all of the bosonic field operators  \cite{Recamier}. Naderi {\it et al} replaced $\hat{a}$ and $\hat{a}^\dagger$ in the standard JCM by the $f$-deformed operators $\hat{A}$ and  $\hat{A}^\dagger$  and  introduced the two-photon $q$-deformed JCM \cite{Naderi}. Barzangeh {\it et al} investigated the effect of a classical gravitational field on the dynamical behavior of nonlinear atom-field interaction within the framework of the $f$-deformed JCM \cite {Barzanjeh}. Abdel-Aty {\it et al} studied the entropy squeezing of a two-level atom in a Kerr medium and examined the influence of the nonlinear interaction of the Kerr medium on the quantum information and entropy squeezing \cite {Abdel-Aty}. Cordero and R\'{e}camier considered the Jaynes-Cummings Hamiltonian with deformed operators of the field and additional Kerr term which is introduced by means of a purely algebraic method \cite{Cordero}. Recently, Faghihi {\it et al} investigated the entanglement dynamics of the nonlinear interaction between a three-level atom (in a $\Lambda$ configuration) and a two-mode cavity field in the presence of a cross-Kerr medium and its deformed counterpart \cite{Honarasa}, intensity-dependent atom-field coupling and the detuning parameters \cite{faghihi2,faghihi3}. Also, the authors have studied a three-level atom in motion which interacts with a single-mode field in an optical cavity in an intensity-dependent coupling regime \cite{faghihi4}. The effects of the mean photon number, detuning, Kerr-like medium and various of the intensity-dependent coupling functional on the degree of entanglement in the interaction between $\Lambda$-type three-level atom with a two-mode field have been studied in \cite{Hanoura}.

Abdalla {\it et al} considered the interaction of a two-level atom with a single-mode multi-photon field in the medium consisting the Stark shift and Kerr medium effects, with the coupling term which is assumed to be a function of time, but still linear in the intensity of light \cite{Abdalla M S}.
Moreover, it is (partially) nonlinear  only due to the presence of the Kerr medium. They investigated atomic inversion and entropy squeezing and showed that, the existence of the time-dependent coupling parameter leads to a time delaying in the interaction which is twice the delay time for time-independent case.
We aimed the present paper to nonlinearize the latter atom-field system which is considered in \cite{Abdalla M S}.
Precisely speaking, the interaction will be occured in a time-dependent and at the same time nonlinear manner between a two-level atom and a nonlinear single-mode field for $k$-photon transitions in the presence of the Stark shift effect and  Kerr medium, both of which are considered to be $f$-deformed. As is clear, in this way, all terms in the Hamiltonian will behave in a nonlinear regime by entering a nonlinearity function $f(n)$, which is generally a well-defined function of the intensity of light. Fortunately, the complicated considered system can be solved analytically and therefore, we will be able to evaluate the time evolution of some of the physical properties such as atomic inversion and entropy squeezing. Since, the exact solution depends on the initial atom-field state, we take the atom to be in its exited state, but the field is considered to be in coherent state, squeezed state and thermal state. Although, our proposal may work well for any arbitrary nonlinear physical system with known $f(n)$, the effect of various parameters and different initial field states will be investigated in detail by considering a particular nonlinearity function.

The paper is organized in the following way: in section 2 we introduce the  interaction Hamiltonian of our considered system in the full nonlinear regime of $k$-photon JCM  and then by solving the corresponding Schr\"{o}dinger equation, the probability amplitudes at any time $t$ for the whole system with arbitrary initial field state have been obtained. In sections 3 and 4 we investigate temporal evolution of atomic inversion and entropy squeezing, respectively. Section 5 deals with presenting our numerical results for atomic inversion and  entropy squeezing versus the scaled time for single- and two-photon transitions and so we discuss about the effects of the Kerr medium, Stark shift, detuning and intensity-dependent coupling on the evolution of mentioned properties. Also, the results of the effects of three- and four-photon transitions on the time evolution of the atomic inversion and entropy squeezing are given in section 6. Finally, we give a summary and conclusion in section 7.
 \section{The $k$-photon JCM: full nonlinear regime}
 The  Hamiltonian of a two-level atom interacting with a quantized field  by the standard JCM in the dipole and the rotating wave approximations can be simply written as ($\hbar=1$), $\hat{H}=\nu \hat{a}^\dagger \hat{a} +\frac{\omega}{2}\hat{\sigma}_{z}+\lambda ( \hat{a}^\dagger\hat{\sigma}_{-}+\hat{a}\hat{\sigma}_{+})$, where $\hat{\sigma}_{z}$ and  $\hat{\sigma}_{\pm}$ are the Pauli operators, $\hat{a}$ and $\hat{a}^\dagger$ are the bosonic annihilation and creation operators, $\nu$ is the frequency of the field, $\omega$ is the transition frequency between the excited and ground states of the atom and $\lambda$ is the constant coupling between atom and field. By a few steps of generalizing standard JCM, the time dependent single-mode $k$-photon JCM in the presence of linear Stark shift and Kerr medium with the time dependent coupling has been studied by the Hamiltonian \cite{Abdalla M S}
 \begin{eqnarray}\label{2}
\hat{H}(t)&=&\nu \hat{a}^\dagger \hat{a}+\frac{\omega}{2}\hat{\sigma}_{z}+  \hat{a}^\dagger \hat{a}(\beta_{1}|g\rangle\langle g|+\beta_{2}|e\rangle\langle e|)\nonumber\\ &+& \chi \hat{a}^{\dagger 2} \hat{a}^{2}+\lambda(t)(\hat{a}^{\dagger k}\hat{\sigma}_{-}+\hat{a}^{k}\hat{\sigma}_{+}),
\end{eqnarray}
where $\beta_{1}$ and $\beta_{2}$ are the effective Stark coefficients, $\chi$ denotes the third-order susceptibility of Kerr medium and $\lambda(t)$ is the time-dependent coupling parameter.
The third term in Hamiltonian (\ref{2}) indicates to the linear (in $a^\dag a$) Stark shift effect, which is arisen from the virtual transition to the intermediate level \cite{Puri,Ahmad,Obada}, and can  exist for two-photon transition, i.e.,  $k=2$ \cite{Puri}. So, for instance,  the authors of Refs. \cite{Abdalla M S,Liao2} used $\delta_{k,2}$ besides the Stark shift term in their  Hamiltonian. In addition, it should be emphasized that, in the Hamiltonian (\ref{2}), whenever  $k\neq2$ one has to set $\beta_{1}=0=\beta_{2}$ \cite{Liao2}. Altogether, it is worth to mention that, the (nonlinear) Stark shift can also exist for the cases with $k>2$ \cite{Ahmad}. In this latter case, the Hamiltonian (\ref{2}) changes to the following form
\begin{eqnarray}\label{2,1}
\hat{H}(t)&=&\nu \hat{a}^\dagger \hat{a}+\frac{\omega}{2}\hat{\sigma}_{z}+  \hat{a}^{\dagger k} \hat{a}^{k}(\beta_{1}|g\rangle\langle g|+\beta_{2}|e\rangle\langle e|) \nonumber\\ &+& \chi \hat{a}^{\dagger 2} \hat{a}^{2}+\lambda(t)(\hat{a}^{\dagger 2k}\hat{\sigma}_{-}+\hat{a}^{2k}\hat{\sigma}_{+}).
\end{eqnarray}
The Hamiltonian (\ref{2,1}) for $k=1$ is equal to the Hamiltonian (\ref{2}) for $k=2$ (linear Stark shift can be occured). Altogether, in the continuation of the paper, by following the path of \cite{Abdalla M S}, we will generalize the Hamiltonian (\ref{2}), which is performed for linear Stark shift effect,  in order to be able to compare our results with the presented results in \cite{Abdalla M S}. Anyway, by defining the detuning parameter $\Delta=\omega-k\nu$, the Hamiltonian (\ref{2}) can be rewritten in the form
 \begin{eqnarray}\label{3}
\hat{H}(t)&=&\nu( \hat{a}^\dagger \hat{a}+\frac{k}{2}\hat{\sigma}_{z})+\frac{\Delta}{2}\hat{\sigma}_{z}+  \hat{a}^\dagger \hat{a}(\beta_{1}|g\rangle\langle g|+\beta_{2}|e\rangle\langle e|)\nonumber\\ &+& \chi \hat{a}^{\dagger 2} \hat{a}^{2}+\lambda(t)(\hat{a}^{\dagger k}\hat{\sigma}_{-}+\hat{a}^{k}\hat{\sigma}_{+}).
\end{eqnarray}
The aim of this paper is to generalize all terms of the Hamiltonian (\ref{3}) via the well-known nonlinear coherent state approach \cite{Manko,Vogel2}. By the notion of the nonlinearity we mean that, we intend to enter the $f$-deformation function in all possible terms, i.e. we will replace all $\hat{a}$ and $\hat{a}^\dagger$ respectively by $\hat{A}=\hat{a} f(\hat{n})$ and $\hat{A}^\dagger=f(\hat{n})\hat{a}^\dagger$ where $f(\hat{n})$ is a function of the number operator (intensity of light). By performing the mentioned procedure, the full nonlinear single-mode $k$-photon time-dependent  JCM in the presence of effective Stark shift and Kerr medium  can be written in the following manner
\begin{eqnarray}\label{5}
\hat{H}(t)&=&\nu (\hat{A}^\dagger \hat{A}+\frac{k}{2}\hat{\sigma}_{z})+\frac{\Delta}{2}\hat{\sigma}_{z}+ \hat{A}^\dagger \hat{A}(\beta_{1}|g\rangle\langle g|+\beta_{2}|e\rangle\langle e|)\nonumber\\&+&\chi \hat{A}^{\dagger 2}\hat{A}^{2}+\lambda(t)( \hat{A}^{\dagger k}\hat{\sigma}_{-}+\hat{A}^{k}\hat{\sigma}_{+}),
\end{eqnarray}
In this respect, a few words seems to be necessary about our present work. It may be recognized that, starting with the nonlinear Hamiltonian describing the interaction between a three-level atom and a single-mode $f$-deformed cavity field (without the Stark shift) and following the path of Refs. \cite{Puri,Ahmad}, the same equations of motion for the three levels of the atom will be achieved. Therefore, one can conclude that, replacing $a, a^\dag$ with $A, A^\dag$
does not change the final results of the above Refs. By these explanations, we would like to emphasize that, the Stark shift should exist in the
 generalized form of the Hamiltonian (\ref{5}), too. In other words, the Stark shift coefficients are now linear in terms of $\hat{A}^\dag \hat{A}$,
 i.e., the field part of the Hamiltonian (\ref{5}).
 So, in Hamiltonian (\ref{5}) (similar to (\ref{2}) and (\ref{3})), the linear (in terms of $\hat{A}^\dag \hat{A}$) Stark shift can exist for the case $k=2$.
 And whenever $k\neq2$ one should set $\beta_{1}=0=\beta_{2}$.
To see what we have really done explicitly, it can be easily seen that
\begin{eqnarray}\label{6}
\hat{H}(t)&=&\nu( \hat{a}^\dagger \hat{a} f^{2}(\hat{n})+\frac{k}{2}\hat{\sigma}_{z})+\frac{\Delta}{2} \hat{\sigma}_{z}+\hat{a}^\dagger \hat{a} f^{2}(\hat{n})(\beta_{1}|g\rangle\langle g|+\beta_{2}|e\rangle\langle e|) \nonumber\\ &+&\chi f^{2} (\hat{n}) f^{2}(\hat{n}-1)\hat{a}^{\dagger2} \hat{a}^{2}+\lambda(t)\left(\frac{\left[ f(\hat{n})\right] !}{\left[ f(\hat{n}-k)\right] !} \hat{a}^{\dagger k}\hat{\sigma}_{-} +\hat{a}^{ k}\frac{\left[ f(\hat{n})\right] !}{\left[ f(\hat{n}-k)\right] !} \hat{\sigma}_{+}\right),
\end{eqnarray}
where $\hat{n}=\hat{a}^\dagger \hat{a}$ and $\left[ f(\hat{n})\right]! \doteq f(\hat{n})f(\hat{n}-1).....f(1)$ with $\left[ f(0)\right] !\doteq1$.
As is clear from (\ref{6}) in comparison with previous Hamiltonian (\ref{3}) which is considered in  \cite{Abdalla M S}, we have in fact made the transformations, $ \nu \rightarrow \nu f^{2}(\hat n), \beta_{1(2)}\rightarrow\beta_{1(2)}f^{2}(\hat n), \chi\rightarrow\chi f^{2}(\hat{n})f^{2}(\hat{n}-1)$ and $\lambda(t) \rightarrow \lambda(t) \frac{[f(\hat{n}]!}{[f(\hat{n}-k]!}.$
It is seen that, the field frequency, Stark shifts, third-order susceptibility and time-dependent parameter are all evolved from $c$-numbers to operator-valued functions (intensity-dependent parameters) \cite{Buz^ek,Faghihi,Singh,Manko,Honarasa}.

The time-dependent $\lambda$-parameter makes the whole Hamiltonian to be time-dependent. Different forms may be chosen for $\lambda(t)$. In this paper we will select $\lambda(t)=\gamma\cos(\mu t)$, where $\gamma$ and $\mu$ are arbitrary constants. Following the probability amplitude approach \cite{Scully}, we assume that,  the wave function of the atom-field can be expressed as \cite{Wolfang P}
 \begin{equation}\label{7}
 \hspace{-1in}|\psi(t)\rangle=\sum_{n} \exp[-i\nu(\hat{n}f^{2}(\hat{n})+\frac{k}{2}\hat{\sigma}_{z})](c_{n,e}(t)|n,e\rangle+c_{n+k,g}(t)|n+k,g\rangle),
 \end{equation}
 where $|n,e\rangle$ and $|n+k,g\rangle$ are the states in which the atom is in exited and ground state and the field has $n$ and $n+k$ photons, respectively. Setting the wave function (\ref{7}) in the time dependent Schr\"{o}dinger equation, $i\hbar\frac{\partial}{\partial t}|\psi(t)\rangle=\hat{H}|\psi(t)\rangle$, we obtain the following coupled equations for $c _{n,e}(t)$ and $c_{n+k,g}(t)$:
  \begin{eqnarray}\label{9}
i\frac{d c_{n,e}}{dt}&=&R_{1}c_{n,e}(t)+\alpha_{n} \cos(\mu t)c_{n+k,g}(t), \nonumber \\i\frac{d c_{n+k,g}}{dt}&=&R_{2}c_{n+k,g}(t)+\alpha_{n} \cos(\mu t)c_{n,e}(t),
\end{eqnarray}
where $R_{1}$, $R_{2}$ and $\alpha_{n}$ are defined as follows:
 \begin{eqnarray}\label{10}
R_{1}&=&\frac{\Delta}{2}+n f^{2}(n)\beta_{2}+\chi n(n-1)f^{2}(n)f^{2}(n-1),\nonumber\\ R_{2}&=&- \frac{\Delta}{2}+(n+k) f^{2}(n+k)\beta_{1}\nonumber\\ &+&\chi (n+k)(n+k-1)f^{2}(n+k)f^{2}(n+k-1), \nonumber\\\alpha_{n}&=&\gamma\frac{\left[ f(n+k)\right] !}{\left[ f(n)\right] !}\sqrt{\frac{(n+k)!}{n!}}.
\end{eqnarray}
The fast frequency dependence of $c _{n,e}(t)$ and $c_{n+k,g}(t)$ can be removed by transforming them to the slowly varying functions $X(t)$ and $Y(t)$ as
 \begin{equation}\label{11}
X(t)=c_{n,e}(t) \exp(iR_{1}t), \hspace{1.5cm} Y(t)=c_{n+k,g}(t) \exp(iR_{2}t).
 \end{equation}
On using (\ref{11}) in equation (\ref{9}) we obtain
 \begin{eqnarray}\label{12}
\frac{d X}{dt}&=&-i\frac{\alpha_{n}}{2}(\ e^{i(\mu+R_{n})t}+\ e^{-i(\mu-R_{n})t})Y,\nonumber\\\frac{d Y}{dt}&=&-i\frac{\alpha_{n}}{2}(\ e^{i(\mu-R_{n})t}+\ e^{-i(\mu+R_{n})t})X,
\end{eqnarray}
where
  \begin{eqnarray}\label{13}
 R_{n} &=& R_{1}-R_{2}\nonumber\\
 &=& \Delta +\chi[n(n-1)f^{2}(n)f^{2}(n-1)-(n+k)(n+k-1)f^{2}(n+k)f^{2}(n+k-1)]\nonumber\\
 &+& [nf^{2}(n)\beta_{2}-(n+k)f^{2}(n+k)\beta_{1}].
 \end{eqnarray}
 The coupled differential equations in (\ref{12}) consist of two terms; the term in the form $e^{i(\mu - R_{n})t}$ ($e^{i(\mu+R_{n})t}$) describes the process that, energy is conserved (nonconserved). So we neglect the terms corresponding to nonconserving energy (in the rotating wave approximation). Under this condition, equations in (\ref{12}) change to
 \begin{eqnarray}\label{14}
\frac{dX}{dt}&=&-i\frac{\alpha_{n}}{2}\ e^{-i(\mu-R_{n})t} Y, \nonumber\\\frac{d Y}{dt}&=&-i\frac{\alpha_{n}}{2}\ e^{i(\mu-R_{n})t}X.
\end{eqnarray}
By solving the above coupling equations, we obtain
\begin{eqnarray}\label{15}
c_{n,e}(t)&=& \left \lbrace c_{n,e}(0)\left (\cos(\Omega_{n}t)-i(R_{n}-\mu)\frac{\sin( \Omega_{n}t)}{2 \Omega_{n}}\right) -i\frac{\alpha_{n}}{2\Omega_{n}}\sin(\Omega_{n}t)c_{n+k,g}(0)\right\rbrace \nonumber\\
&\times& \exp [-i(\varphi_{n}+\mu/2)t],
\end{eqnarray}
\begin{eqnarray}\label{16}
c_{n+k,g}(t) &=& \left \lbrace c_{n+k,g}(0)\left (\cos(\Omega_{n}t)+i(R_{n}-\mu)\frac{\sin(\Omega_{n}t)}{2\Omega_{n}}\right)-i\frac{\alpha_{n}}{2\Omega_{n}}\sin(\Omega_{n}t)c_{n,e}(0)\right\rbrace \nonumber\\
&\times& \exp[-i(\varphi_{n}-\mu/2)t],
\end{eqnarray}
where
\begin{eqnarray}\label{17}
\varphi_{n}&=&\frac{\chi}{2}[n(n-1)f^{2}(n)f^{2}(n-1)+(n+k)(n+k-1)f^{2}(n+k)f^{2}(n+k-1)]\nonumber\\
 &+&\frac{1}{2}[nf^{2}(n)\beta_{2}+(n+k)f^{2}(n+k)\beta_{1}].
\end{eqnarray}
and $\Omega_{n}=\frac{1}{2}\sqrt{(R_{n}-\mu)^{2}+\alpha_{n}^{2}}$ is the generalized Rabi frequency (note that $\alpha_{n}$ and $R_{n}$ are defined respectively in (\ref{10}) and (\ref{13})). It ought to be  mentioned that, in equations (\ref{10}), (\ref{13}) and (\ref{17}), the values $\beta_{1}=0=\beta_{2}$ should be set for the case $k\neq2$.

In the above equations, $c_{n,e}(0)$ and  $c_{n+k,g}(0)$ may be determined with the initial states of atom and field. In this work we suppose that, the atom is initially in the excited state ($|e\rangle  $), however, the cavity field is considered to be initially in different states such as coherent state, squeezed state and thermal state which can be defined by their associated density operators as
\begin{equation}\label{18}
\rho_{CS}(0)= |\alpha\rangle \langle\alpha|=e^{-\langle n \rangle}\sum_{n,m=0}^{\infty}\frac{\alpha^{m}\alpha^{\ast n}}{\sqrt{m! n!}}|m\rangle\langle n|
\end{equation}
\begin{equation}\label{19}
\rho_{SS}(0)=|\xi\rangle \langle\xi|=\sum_{n,m=0}^{\infty}\frac{(\tanh r)^{n+m} \sqrt{(2m)!(2n)!}}{(2^{n}n!2^{m}m!)\cosh r} |2m\rangle\langle 2n|,
\end{equation}
\begin{equation}\label{20}
\rho_{TS}(0)=\sum_{n=0}^{\infty}\frac{\langle n\rangle^{n}}{(1+\langle n\rangle)^{n+1}}|n\rangle\langle n|.
\end{equation}
Therefore, by using each of the above initial field conditions we can find the explicit form of the solution of time dependent Schr\"{o}dinger equation. This enables one to analyze interesting properties such as atomic inversion and entropy squeezing, which will be done in the following sections.

It can clearly be seen that, setting $f(n)=1$ in the relations (\ref{15}), (\ref{16}) recovers the results of Ref. \cite{Abdalla M S}. As another important point, it is worth mentioning that,  choosing different nonlinearity functions leads to different Hamiltonian systems and so, different physical results may be achieved. Altogether, in the continuation of this paper,  we select particularly the intensity dependent coupling as $f(n)=\sqrt{n}$. This function is a favorite function for the authors who have worked in the nonlinear regime of atom-field interaction (see for instance \cite{Singh}, \cite{Huang}). In particular, Fink {\it et al} have explored a natural way that, this nonlinearity function will be appeared in physical systems \cite{Fink}.

\section{Atomic inversion}
The atomic inversion measures the difference in the populations of the two levels of the atom and plays a fundamental role in laser theory \cite{Wolfang P}. After determining $c_{n,e}(t)$ and $c_{n+k,g}(t)$ for the  initial field states in (\ref{18}), (\ref{19}) and (\ref{20}), we can investigate this quantity which is given by
 \begin{equation} \label{21}
W(t)=\sum_{n=0}^{\infty}(|c_{n,e}(t)|^{2}-|c_{n+k,g}(t)|^{2}).
\end{equation}
By inserting the equations (\ref{15}) and (\ref{16}) in (\ref{21}) for an arbitrary initial field state, we obtain
 \begin{equation}\label{22}
 W(t)=\sum_{n=0}^{\infty}\rho_{nn}(0)\left(\cos(2\Omega_{n} t)+(R_{n}-\mu)^{2}\frac{\sin^{2}(\Omega_{n} t)}{2 \Omega_{n}^{2}}\right),
\end{equation}
where $\rho_{nn}(0)=|c_{n}(0)|^{2}$. For the mentioned initial field states in (\ref{18}), (\ref{19}) and (\ref{20}) and one has:
\begin{equation}\label{221}
\rho^{CS}_{n,n}(0)= |c_n^{CS}(0)|^{2}= e^{-\langle n \rangle}\frac{\langle n \rangle ^{2 n}}{ n!},
\end{equation}
\begin{equation}\label{222}
\rho^{SS}_{2n,2n}(0)= |c_{2n}^{SS}(0)|^{2}=\frac{\langle n\rangle^{n} (2n)!}{(2^{n}n!)^2(1+\langle n\rangle)^{n+1/2}},\hspace{1cm}        \rho^{SS}_{2n+1,2n+1}(0)= |c_{2n+1}^{SS}(0)|^{2}=0
\end{equation}
\begin{equation}\label{223}
\rho^{TS}_{n,n}(0)=\frac{\langle n\rangle^n }{{(1+\langle n\rangle)^{n+1}}},
\end{equation}
where $\langle n\rangle$ for each of these states is given by
\begin{equation}\label{224}
\langle n\rangle_{CS}=|\alpha|^{2},\hspace{0.5cm} \langle n\rangle_{SS}=\sinh^{2}(r),\hspace{0.5cm}\langle n\rangle_{TS}=\frac{1}{e^{\hbar\nu/k_{B}T}-1}.
\end{equation}\label{22,4}
 From equation (\ref{22}) we can discuss the temporal evolution of the atomic inversion for different initial field situations. This will be presented in section 5 in detail.
 \section{Entropy squeezing }

 For a two-level atom, characterized by the Pauli operators $\sigma_{x}$, $\sigma_{y}$ and $\sigma_{z}$, the uncertainty relation for the information entropy is defined as follows \cite{Fang}
\begin{equation}\label{23}
\delta H(\sigma_{x})\delta H(\sigma_{y})\geq\frac{4}{\delta H(\sigma_{z})},\hspace{2cm}\delta H(\sigma_{\alpha})=\exp[H(\sigma_{\alpha})],
\end{equation}
 where $ H(\sigma_{\alpha})$, as the information entropy of the operator $ \sigma_{\alpha}(\alpha=x,y,z)$, is given by
 \begin{equation}\label{24}
 H(\sigma_{\alpha})=-\sum_{i=1}^{2}P_{i}(\sigma_{\alpha})\ln P_{i}(\sigma_{\alpha}).
 \end{equation}
Since for a two-level atom, the Pauli operators have two eigenvalues, one may expect that, $P_{i}(\sigma_{\alpha})$ denotes the
probability distribution of two possible outcomes of measurements of the  operator $\sigma_{\alpha}$. Henceforth, it is defined as follows
 \begin{equation}\label{25}
P_i (\sigma_{\alpha})= \langle \psi_{\alpha_i}| \rho|\psi_{\alpha_i} \rangle ,
 \end{equation}
where $\rho$ is the density operator of the system and $|\psi_{\alpha_i} \rangle$ is the eigenstate of the Pauli operators, i.e.,
\begin{equation}\label{26}
\sigma_{\alpha} | \psi_{\alpha_i} \rangle  = \eta_{\alpha_i}| \psi_{\alpha_i} \rangle,\hspace{2cm}\alpha=x,y,z,    \hspace{0.25cm}  \hspace{.25cm} i=1,2.
\end{equation}
From equation (\ref{23}) the components $ \sigma_{\alpha}(\alpha=x,y)$ are said to be squeezed, if the information entropy  $H(\sigma_{\alpha})$ of $\sigma_{\alpha}$ satisfies the inequality
\begin{equation}\label{27}
 E(\sigma_{\alpha})=\delta H(\sigma_{\alpha})-\frac{2}{\sqrt{\delta H(\sigma_{z})}}<0,\hspace{2cm}\alpha=x \hspace{.25cm} or\hspace{.25cm} y.
\end{equation}
By using the equations (\ref{24}) and (\ref{25}) for the information entropies of the atomic operators $\sigma_{x}$, $\sigma_{y}$ and $\sigma_{z}$ we finally arrive at
\begin{eqnarray}\label{29}
H(\sigma_{x})=&-&\left[ \frac{1}{2}+Re(\rho_{ge}(t))\right] \ln\left[ \frac{1}{2}+Re(\rho_{ge}(t))\right]\nonumber\\&-&\left[ \frac{1}{2}-Re(\rho_{ge}(t))\right] \ln\left[ \frac{1}{2}-Re(\rho_{ge}(t))\right],
\end{eqnarray}
\begin{eqnarray}\label{30}
H(\sigma_{y})=&-&\left[ \frac{1}{2}+Im(\rho_{ge}(t))\right] \ln\left[ \frac{1}{2}+Im(\rho_{ge}(t))\right]\nonumber\\&-&\left[ \frac{1}{2}-Im(\rho_{ge}(t))\right] \ln\left[ \frac{1}{2}-Im(\rho_{ge}(t))\right],
\end{eqnarray}
\begin{eqnarray}\label{31}
H(\sigma_{z})=-\rho_{ee}(t) \ln\rho_{ee}(t)-\rho_{gg}(t)\ln\rho_{gg}(t).
\end{eqnarray}
By using the form of the wave function (\ref{7}), the density operator of the entire atom-field system at any time $t$ is given by
\begin{eqnarray}\label{28}
\hspace{-1cm}\rho_{\mathrm{atom-field}} =\sum_{n=0}^{\infty}\sum_{m=0}^{\infty} \lbrace c_{n,e}(t)c_{m,e}^{*}(t)|n,e\rangle \langle e,m|+ c_{n+k,g}(t)c_{m+k,g}^{*}(t)|n+k,g\rangle \langle g,m+k|\nonumber\\ \hspace*{-.41in} +c_{n,e}(t)c_{m+k,g}^{*}(t)|n,e\rangle \langle g,m+k|+ c_{n+k,g}(t)c_{m,e}^{*}(t)|n+k,g\rangle \langle e,m|\rbrace.
\end{eqnarray}
So, the necessary matrix elements of the reduced density operator in (\ref{29})-(\ref{31}) may be given in the following form
\begin{eqnarray}\label{32}
\rho_{ee}(t)=\sum_{n=0}^{\infty}|c_{n,e}(t)|^{2},
\end{eqnarray}
\begin{eqnarray}\label{33}
\rho_{eg}(t)=\sum_{n=0}^{\infty}c_{n+k,e}(t)c_{n+k,g}^{*}(t)=\rho_{ge}^{*}(t),
\end{eqnarray}
\begin{eqnarray}\label{34}
\rho_{gg}(t)=\sum_{n=0}^{\infty}|c_{n+k,g}(t)|^{2}.
\end{eqnarray}
By employing the above equations, we can study the temporal evolution of the entropy squeezing in terms of the variables $\sigma_{x}$ and $\sigma_{y}$, which will be done in the next section.
By using equation (\ref{22}) and replacing $\rho_{nn}(0)$ for different initial field states (coherent, squeezed and thermal states from (\ref {221}), (\ref{222}) and (\ref{223})), we can investigate the effects of the initial field state on the variation of the atomic inversion. Prior to everything it is necessary to select a particular nonlinearity function. As we mentioned previously, in this paper we choose $f(n)=\sqrt{n}$.
In all figures which are related to the atomic inversion $W(t)$, the left and the right plots respectively corresponds to the linear and nonlinear function.
All figures are drawn with particular values of $\mu=0.1$, $\langle n \rangle=25$. Other used parameters are denoted in the related figure captions distinctly.
Figure 1 shows the temporal evolution of the atomic inversion in terms of the scaled time, for the functions $f(n)=\sqrt{n}$ and also $f(n)=1$ taking the  ``coherent state''  in (\ref{18}) as the initial field state.
Figures 1(a) and 1(b) show the variation of the atomic inversion without Kerr and Stark effects.
The collapse and revival phenomena exist in these figures, but there is an increase in the number of fluctuations with regular behavior for the deformed case. Also, the amplitude of the fluctuation for this case is increased relative to $f(n)=1$. In other words, while we have partial revivals in the linear case, nearly complete revivals occur in the nonlinear regime.
To examine the effect of the Kerr medium on the behavior of the population inversion, figures 1(c) and 1(d) are plotted. Figure 1(d) which corresponds to $f(n)=\sqrt{n}$ shows a chaotic behavior of $W(t)$ around $0.99$, such that  the amplitude of the fluctuations between maxima and minima of $W(t)$ are very small. Figure 1(c) indicates that, in the presence of the Kerr effect for the case $f(n)=1$, the result is very similar to figure 1(a). Altogether, if we use the value of $\chi$ larger (up to 0.03 \cite{Abdalla M S}) the Kerr effect will be visible.
The effect of the Stark shift (in the presence of Kerr medium) can be seen for linear and nonlinear functions in figures 1(e) and 1(f). From figure 1(f), we observe that, the Stark shift increases the amplitude of the fluctuations as compared with figure 1(d). Also, this figure shows a chaotic behavior for $W(t)$ in the nonlinear regime.
Figure 1(e) shows the effect of the Stark shift (in the presence of Kerr medium) on the time variation of  $W(t)$ for $f(n)=1$. One can see that, the temporal evolution of the atomic inversion reveals several revivals in the presence of both the Stark and Kerr effects.
Comparing figures 1(e) and 1(b) leads us to conclude that, the effect of the considered nonlinearity function (without the Kerr and Stark effects) is nearly equivalent to the Kerr and Stark effects in the linear case.
The effect of the detuning parameter $\Delta$ (defined as $\omega-k\nu=\Delta$), in the presence Kerr and Stark effects has been shown in figures 1(g) and 1(h). Comparing figure 1(g) with figure 1(e) indicates that the extremes of $W(t)$ (in the revivals) are regularly decreases for the linear system in our plotted figure (figure 1(g)). Altogether, $\Delta$ has a negligible effect in the presence of nonlinearity function (figure 1(h)).
We have plotted figure 2 taking into account the initial field as ``squeezed state" using (\ref{19}). Figures 2(a) and 2(b) show the time evolution of the atomic inversion for the linear and nonlinear function, in the absence of both Kerr and Stark effects. We can see from figure 2(a) that, $W(t)$ oscillates rapidly for the  case $f(n)=1$, while in the presence of nonlinearity the behaviour of $W(t)$ is periodic (figure 2(b)). Figures 2(c) and 2(d) demonstrate the Kerr medium effect on the variation of $W(t)$ and in figures 2(e) and 2(f), we added the Stark shift, too. By comparison the figures 2(a) and 2(c), one finds that, the behavior of $W(t)$ for $f(n)=1$, with and without Kerr effect, are almost the same, however, for the linear case and in the presence of the Stark effect (figure 2(e)) the behavior of $W(t)$ is irregular. We examined the time evolution of the atomic inversion for the nonlinear case with different parameters in the right plots of figure 2. We can see a regular behavior for $W(t)$ in the absence of the Kerr medium, Stark effect and detuning (figure 2(b)).
But, the behavior of $W(t)$ in the figures 2(d), 2(f) and 2(h) is generally irregular.
Figures 2(g) and 2(h) show the effect of the detuning parameter for linear and nonlinear function. We can see partial revivals in the presence of detuning.
In figure 3, we assumed that the initial field is ``thermal state" which is defined in (\ref{20}) and again  the effects of the Kerr medium, Stark shift and detuning are investigated on the behavior of $W(t)$. The evolution of the atomic inversion is shown for nonlinear and linear regimes, in the right and left plots of this figure, respectively.
We have shown the effect of Kerr medium in figure 3(c) for linear function, where one can see that, $W(t)$ is not so sensitive to the Kerr effect.
While in figure 3(b), the presence of nonlinearity without both Kerr and Stark effects, allows the (partial) collapses and revivals to be observed, the Kerr medium and Stark shift destroy the latter phenomena.
The above result seems to be in contrast to the linear case, i.e., the presence of Kerr and Stark shift effects can appear the (partial) collapses and revivals apparently.
We observe that the variation of $W(t)$ with $\Delta\neq0$ in the presence of the Kerr and Stark effects for linear and nonlinear function in figures 3(g) and 3(h), respectively. As in the previous states considered in this paper, unlike some changes in the numerical results, no qualitatively change can be observed. Altogether, generally in all three states discussed above, linear case is more sensitive to detuning parameter in comparison with nonlinear case.\\
In this part of the present section, we will analyze the temporal evolution of the entropy squeezing for different initial field states using the analytical results of section 4.
 We will deal with nonlinear case  with deformation function $f(n)=\sqrt{n}$ only. All figures are drawn with particular value of $\mu=0.1$. Other used parameters are denoted in the related figure captions distinctly.
Figures 4(a) and 4(b) display the time evolution of the entropy squeezing factors $E(\sigma_{x})$ and $E(\sigma_{y})$  if one concerns with the initial field as a ``coherent state" in (\ref {18}). It is obvious from these figures that, there exists entropy squeezing in $\sigma_{x}$ and $\sigma_{y}$ at some intervals of time.
In figures 4(c) and 4(d), we examine the influence of the Kerr effect on the evolution of the entropy squeezing  for the variables $\sigma_{x}$ and $\sigma_{y}$ with the chosen parameters, respectively.
It is clear from these figures that, there is no squeezing in $\sigma _{x}$ and $\sigma_{y}$. Also, paying attention to figures 4(e) and 4(f) which are plotted in the presence of both Kerr and Stark effects, no squeezing can be seen in  $\sigma_{x}$ and $\sigma_{y}$.
To study the effect of the initial mean photon number on the  behavior of the entropy squeezing (with Kerr effect) figures 4(g) and 4(h) are plotted.
 As is shown, by decrement the mean value of $ \langle n \rangle $ from 25 to 1, the entropy squeezing for the variables $\sigma_{x}$ and $\sigma_{y}$  will be appeared in certain time ranges. The time evolution of the squeezing parameters $E(\sigma_{x})$ and $E(\sigma_{y})$ are shown in figure 5, for the field initially being in the ``squeezed state" in (\ref{19}). Specifically, in figures 5(a) and 5(b), the behavior of the  squeezing  $E(\sigma_{x})$ and $E(\sigma_{y})$ as a function of the scaled time in the absence of the Kerr and Stark effects have been shown. We see from these figures that, both $E(\sigma_{x})$ and $E(\sigma_{y})$ possess squeezing in the variables $\sigma_{x}$ and $\sigma_{y}$  when $ \langle n \rangle =1$. It should be noticed that, according to our further calculations (not shown here) in this case (without Kerr and Stark effect),  squeezing may be seen in the components $\sigma_{x}$ and $\sigma_{y}$ for $ \langle n \rangle < 4$.
To investigate the effect of the Kerr medium, we have depicted the entropy squeezing $E(\sigma_{x})$ and $E(\sigma_{y})$ in terms of the scaled time in figures 5(c) and 5(d). $E(\sigma_{x})$ and $E(\sigma_{y})$ predict squeezing in the variables $\sigma_{x}$ and $\sigma_{y}$ on short time periods discontinuously.
A comparison of the figures 5(a), 5(b), 5(c) and 5(d) with  similar figures for coherent state (figures 4(a), 4(b), 4(c) and 4(d)) shows that, while for the second set of figures the Kerr effect destroys the entropy squeezing completely, this is not so for the first set of figures.
We discuss the effects of Stark shift on the time evolution of the squeezing factors in figures 5(e) and 5(f). It is obvious from these figures that, there is no squeezing in the presence of the Kerr and Stark effects.
Figure 6 shows the time evolution of the entropy squeezing factors $E(\sigma_{x})$ and $E(\sigma_{y})$ for the case that, the field is initially prepared in the ``thermal state" in (\ref{20}). Figures 6(a) and 6(b) represent the entropy squeezing  $E(\sigma_{x})$ and $E(\sigma_{y})$ in the absence of the Kerr and Stark effects. As is clear, squeezing in the components $\sigma_{x}$ and $\sigma_{y}$ exists at some  intervals of time, obviously for different time intervals.  Also, the depth of the entropy squeezing for  $\sigma_{x}$ is larger than for $\sigma_{y}$.
We investigated the effect of the Kerr medium in figures 6(c) and 6(d). As is observed, squeezing may be occurred in the entropy squeezing factors $E(\sigma_{x})$ and $E(\sigma_{y})$  in a short range of time.
Finally, we examined the effect of the Stark shifts (when the Kerr effect is also in our consideration) in figures 6(e) and 6(f). In this case, there is no squeezing in $E(\sigma_{x})$ and $E(\sigma_{y})$.

\section{A discussion on the effect of three- and four-photon transitions}
We investigated the influence of one- and two-photon transitions on the temporal behaviour of atomic inversion and entropy squeezing in the previous sections.
In this section, we intend to discuss the effect of three- and four-photon processes on the time  evolution of  mentioned physical quantities  in a general manner.
Obviously, adding all of the numerical results and related figures considering  all quantities which  concern with $k=3, 4$ will make the paper dramatically large.
Therefore, we present our obtained results qualitatively and make a comparison with the previous results for $k=1,2$. Clearly, due the numerous
parameters which are involved in the calculations, one can not reach a sharp result, so, our discussion is restricted to the particular used parameters.
 According to our further calculations for $k=3$ and $k=4$ (not shown here), the following results have been extracted:
\begin{itemize}

  \item  The collapse and revival phenomena exist in a clear manner for three- and four-photon transitions in the linear regime ($f(n)=1$) when the filed is initially in the coherent state. As we observed, by increasing the number of photon transitions, the time interval between subsequent revivals   will be decreased. In addition, the revival times turn shorter when the number of photon transition is increased. This result is in consistence with the outcome results of Ref. \cite{Kang} ([18] of RM).
     Moreover, for $k=3,4$, no clear collapse-revival phenomenon is observed for the atom-field states in the linear regime ($f(n)=1$) which their initial field states are squeezed and thermal states.
  \item The temporal behaviour of atomic inversion for $k=3$ and $k=4$ shows a chaotic behaviour for the nonlinear regime ($f(n)=\sqrt{n}$) in all cases.
 As one may observe, when the initial  field is thermal and coherent states, for the case $k=1$  in the absence of Kerr medium and detuning, the full collapse and revivals are revealed in the evolution of atomic inversion.

    \item Our results show that, Kerr medium has  a negligible effect on the time variation of atomic inversion for all cases with $k=3, 4$. For the detuning parameter
   we observed that for the linear case with $k=3$, it has no critical effect for the coherent and thermal states as initial field, while for the squeezed initial state, the negative values of atomic inversion  are considerably decreased, i.e., it gets positive values in main parts of time. The same statement will be weakly true for the case $k=4$.

  \item In the nonlinear case ($f(n)=\sqrt{n}$), there is no (entropy) squeezing in $\sigma_{x}$ and $\sigma_{y}$ for $k=4$ with different initial field states and also, for $k=3$ with squeezed and thermal states as the initial states of the field. But, for $k=3$ in the absence of Kerr medium, squeezing exists in $\sigma_{y}$ in a very short intervals of times. In this case, there is no squeezing in $\sigma_{x}$, too.
 \end{itemize}
\section{Summary and conclusion}
In this paper, we considered the full nonlinear interaction between a two-level atom with a nonlinear single-mode quantized field for $k$-photon transition in the presence of Kerr medium and Stark shift effect.
Also, we assumed that, the coupling between atom and field is time-dependent as well as intensity-dependent.
To the best of our knowledge, this problem in such a general form has not been considered in the literature up to now.
Fortunately, we could solve the dynamical problem and found  the explicit  form of the state vector of the whole atom-field system analytically.
We have considered the atom to be initially in the exited state and the field in three different possible states (coherent state, squeezed state and thermal state),
and then, the time variation of  atomic inversion and entropy squeezing have been numerically studied and compared with each other. Even though our formalism can be used for any nonlinearity function, we particularly considered  the nonlinearity function $f(n)=\sqrt{n}$ for our further numerical calculations. The obtained results are summarized as follow:\\
1. The temporal evolution of both atomic inversion and entropy squeezing is generally sensitive to the initial field state, but this fact is more visible  for the atomic inversion in comparison with entropy squeezing. \\
2. The behavior of atomic inversion in the presence of nonlinearity (the right plots in all figures) is chaotic, except in some cases, i.e., figures 1(b), 2(b) and 3(b) which are plotted for the cases in which the Kerr and Stark effects are absent and  the initial field state is coherent state, squeezed state and thermal state, respectively. As is observed, the collapse and revival phenomena are revealed in the figures 1(b) and 3(b).\\
3. The complete (partial) collapse and revival, as purely quantum mechanical features, are observed in the left plots of figure 1 (figure 3) corresponds to atomic inversion for initial coherent (thermal) state. \\
4. The detuning parameter has not a critical effect on atomic inversion, unless it causes some minor changes in the extremes of the investigated quantities, either with chaotic or collapse-revival behavior. \\
5. The variation of atomic inversion for different initial field states (coherent state, squeezed state and thermal state) shows that, the time dependent coupling leads to a time delaying which is twice the delay time for the time-independent case. This result is similarly to reported results in \cite{Abdalla M S}.\\
6. There is seen entropy squeezing in $\sigma_{x}$ and $\sigma_{y}$ at some  intervals of time in some cases with different conditions, obviously for different time intervals, such that the uncertainty relation holds.\\
7. The presence of both Stark shift and Kerr medium simultaneously on the entropy squeezing for all cases (different initial field states) prevents the entropy squeezing to be occurred.\\
8. In the absence of Kerr medium, Stark shift, detuning and with constant coupling ($f(n)=1$), with considering the parameters which are used in Ref. \cite{Kang} ([18] of RM), our results recover the numerical results of Ref. \cite{Kang} successfully.  In the absence of the mentioned effects with intensity-dependent but time-independent coupling ($f(n) =\sqrt{n}$, $\mu = 0$), our results are reduced to the ones reported in Ref. \cite{LiN}.\\
9. As previously mentioned, we nonlinearized the atom-field system which has been considered in \cite{Abdalla M S}. Consequently, as is expected, in the linear case ($f(n)=1$) the outcome results are the same as the results in this Ref.\\
Finally, we would like to mention that, our presented formalism has the potential ability to be applied for all well-known nonlinearity functions, such as the center of mass motion of trapped ion \cite{Vogel}, photon-added coherent states \cite{Agarwal,Sivakumar}, deformed photon-added coherent states \cite{Safaeian}, harmonious states \cite{Manko,Sudarshan}, $q$-deformed  coherent states \cite{Naderi,Macfalane,Biedenharn,Chaichian} etc. We have not discussed the effect of the initial field photon number in detail, but it is obvious that, the results may be affected directly by this parameter, as well as all discussed parameters.


%

\end{document}